\documentclass[]{revtex4}
\usepackage{graphicx}
\usepackage{color}
\usepackage{xcolor}
\usepackage{mathrsfs}
\newcommand{\ket}[1]{{|#1\rangle}}
\newcommand{\bra}[1]{{\langle#1|}}
\begin{document}

\title{Contextuality and Quantum Discord
\vspace{-2mm}}

\author{Asma Al-Qasimi}
\email{alqasimi@pas.rochester.edu}
\affiliation{Department of Physics and Astronomy and Center for Coherence and Quantum Optics, University of Rochester, Rochester, NY 14627, USA}

\vspace{-2mm}
\begin{abstract} \noindent

One of the defining differences between classical and quantum systems is how measurements affect them. Here, we compare the approaches of contextuality and quantum discord in capturing quantum correlations in special classes of two-qubit states, demonstrating that although non-discordant states are non-contextual, discordant states are not always contextual.

\end{abstract}

\maketitle

The question of what makes a system classical or quantum is still a subject of debate. We have pieces of the puzzle, many of which were obtained from direct experimental observations; i.e., by interacting directly with reality. However, we are still struggling with connecting everything together. This is partly due to the unintuitive nature of the rules that these observations, which we do not experience in our everyday classical world, impose on us. It is also due to our own desire to \emph{know} everything and not wanting to put up with gaps in our knowledge and imperfections in our predictions,  just like the case of Einstein and his colleagues in the celebrated EPR paper \cite{EPR}, where they were struggling with accepting quantum mechanics as a complete description of reality. Moreover, we are limited in what information we can and cannot access in the laboratory. This can range from the limits imposed by the wavelength of electromagnetic fields used in detecting a \emph{particle} \cite{F1}, such as in Compton scattering, to the multiple and not compatible properties that a system can display all at the same time. In the latter case, we are referring to the dual nature of quantum systems, such as photons and electrons, in behaving like both particles and waves. Bohr famously referred to that peculiarity as \emph{complementarity} \cite{F2}. These concepts were quantified in later years in the Young interference experiment, where the particle-nature of light was associated with the \emph{distinguishability};. i.e., how confident we are that the photon went through one slit and not the other, and the wave-nature of light was associated with the \emph{visibility} of fringes; i.e., how well one distinguishes the dark and light fringes from each other \cite{Zurek, Glauber,Greenberger}. Moreover, in more recent years, this quantified relationship was completed by including the entanglement between the degrees of freedom of the single quantum particle in the picture \cite{Eberly2020}.

We agree that a quantum object can strongly display both particle-like and wave-like properties at the same time, but what quantifications of the quantum property do we need to focus on in order to take advantage of the quantum power in building technologies? The most famous one, which was the main focus of quantum information research for decades, is entanglement, inspired by Schr{\"o}dinger's famous statement that it is ``the characteristic trait of quantum mechanics" \cite{Schrodinger}. Entanglement, simply put, is the inability of the state of a system to be described as a product of its individual components. However, that, on its own, cannot be taken as a signature of \emph{quantumness}. To elaborate, to be in an entangled state, all you need is a system, with multiple degrees of freedom, that can be in a superposition state, and it is a well-known fact that classical systems can exist in a superposition state; for example, multi-mode polarized light beams. That is why it is not surprising that \emph{classical} entanglement exists in classical systems \cite{Sprew1998,QianOL,Aiello,EberlyScripta2016,Asma2020}. To determine if a system that is entangled is quantum or not, that property has to be also paired with non-locality \cite{KarimiBoyd}. That is where the violation of Bell's Inequality needs to be considered \cite{Bell}. To complicate the matter further, classical states have been shown to violate Bell's Inequalities \cite{Sprew1998,EberlyOptica2015}. The next natural question is: \emph{how} can we pose these questions to make sure that the answer we get is telling us whether we have a quantum advantage in the system or not, which we can use to go beyond the classical advantage. One restriction that is proposed in the community is that the entanglement cannot be within the degrees of freedom of a single entity; i.e.; it cannot be a property of a single particle \cite{vanEnk}. The rationale is that the properties of a single quantum particle so much resembles those of a classical wave, so it will confuse us.

In this paper, we will focus on two other attempts to capture quantum correlations that in more recent years have been touted as \emph{the} real signature of \emph{quantumness}: quantum discord and quantum contextuality for two qubits. Both of these quantities are defined by using the peculiar measurement properties in quantum mechanics. Moreover, discord is considered to be \emph{better} at capturing quantum properties than entanglement. For example, the Deterministic Quantum Computing with One Qubit (DQC1) model has been shown to have a quantum advantage even when there is no entanglement, but there is discord, in the system \cite{Knill,Caves,White}. Moreover, if we look at the relationship between entanglement and noise, quantified by entropy \cite{DJ}, and that of discord versus noise \cite{AsmaDiscord}, we see that entanglement cannot exist in a system after it reaches a certain level of disorder. On the other hand, discord can still exist in the system even if its state is very close to the maximally mixed state. It is, therefore, not surprising that discord has been proposed as encompassing entanglement (among other correlations) \cite{Modi}. Similarly, contextuality has been regarded as a \emph{better} property to establish the quantumness of a system, as opposed to entanglement's Bell's inequalities \cite{PolishGang} on their own, especially that the latter, as we mentioned earlier, have been shown to be violated by classical fields.

Let us now elaborate on the \emph{approaches} of discord and contextuality, which we will be using later to shed light on the relationship between them. Discord is a non-local property of quantum systems that is defined based on the idea that measurements disturb quantum systems, but not classical ones, where statistical correlations of a bipartite system is compared before and after a measurement of one subsystem to see if it has changed or not \cite{OllivierZurek}. If it has, that is taken as an indication that the system possesses quantum properties. To add to the confusion of the subject, note that measurements \emph{can} disturb classical systems. Think about unpolarized light passing through a polarizing beamsplitter: It emerges as a polarized beam. That is why we need to tread carefully when exploring this avenue.

In a bipartite quantum system, represented by $\hat{\rho}$, made of qubits A, $\hat{\rho}_A$, and B, $\hat{\rho}_B$, discord is detected or quantified by performing a \emph{local} measurement on B (or A) and observing how that affects a \emph{global} property of the system, namely the mutual information function $I(\hat{\rho})$, which in terms of the von Neumann entropy, $S(\hat{\rho})=- \mbox{Tr}  \{ \hat{\rho} \ \mathrm{log_2} \hat{\rho} \}$, of the density matrices of the system is given by:

\begin{equation}
I(\hat{\rho})=S(\hat{\rho}_A)+S(\hat{\rho}_B)-S(\hat{\rho}).
\label{Irho}
\end{equation}

If the value of this function (\ref{Irho}) changes after the local measurement is performed, then there is discord in the system; otherwise, there is no discord. Of course, since the local measurement on subsystem B is not unique, the final value for discord is obtained after it is minimized over all possible measurements on B (See \cite{Luo,AsmaDiscord} for an explicit description of the procedure for the two-qubit case as well as explicit results for special classes of states). Why should a local measurement on B have that effect? The answer stems from the measurement problem in quantum mechanics: measurements disturb quantum systems, collapsing their states, while they should not disturb classical systems. Moreover, as the measurement that is performed on B is done with complete disregard to the fact that there could be other non-commuting properties of B that are measured at the same time, it is not surprising that system B gets disturbed, affecting its relation to A, and resulting in a non-zero value for discord. This is at the heart of the uncertainty relations in quantum mechanics, which result from such non-commuting relations. That is what makes discord so special, as it captures quantum correlations using the very peculiar properties of quantum systems. If B has no quantum properties; i.e., all the operators representing its properties are classical, then measuring all or a subset of these operators should not affect their measuring outcomes. In this case, the value obtained for discord would be zero. A state of zero discord is, therefore, accepted in the scientific community as a classical state \cite{Modi}.

We know from quantum mechanics that we cannot measure non-commuting operators simultaneously, but we can measure commuting operators at the same time. When we say \emph{cannot}, it does not mean we are \emph{not able to}. For example, we \emph{can} measure position and momentum at the same time, but they will end up \emph{disturbing} each others results. The non-commutativity of these two operators, as we mentioned earlier, is what gives the uncertainty relationship between them. We are not discussing this situation here. Instead, we are focusing on situations with commuting operators when we talk about contextuality and we need to discuss its approach, which we will use to derive the main results of this work. Simply put, the idea behind contextuality is that even \emph{when we are dealing with commuting operators in quantum systems, we cannot assign local measurement results to these operators} \cite{Peres1990}. If we do so, we get inconsistencies \cite{Mermin, PeresBook}. In fact, the Kochen-Specker theorem tells us that all quantum systems of dimension greater than or equal to 3 are contextual \cite{KS}, which implies that 2-qubit systems (dimension = 4) are contextual. We will elaborate on the mathematics of this paragraph when we discuss the results. Following the publication of the Kochen-Specker theorem, other approaches to contextuality followed \cite{revQCont}. However, here we intend to stick to the foundational ideas applied to the two-qubit system: if there is an inconsistency due to assigning local measurement results to a system, then it is \emph{contextual}. Otherwise, it is \emph{non-contextual}. This is also part of the measurement problem in quantum mechanics.

Both discord and contextuality are based on the measurement problem in quantum mechanics. They are based on demonstrating inconsistencies when something local is done to the system, in both cases something to do with measurement. Should they not agree then, when we ask the following question: Are there any quantum properties in the system?

To answer this question, we will start our analysis, which follows the approach in \cite{Peres1990}. Consider the following two sets of commuting operators: $\{ \hat{X} \otimes \hat{X}, \hat{Y} \otimes \hat{Y}  \}$ and $\{ \hat{X} \otimes \hat{Y}, \hat{Y} \otimes \hat{X}  \}$, where $\hat{X}$ and $\hat{Y}$ are the Pauli spin operators $\hat{\sigma}_{x}$ and $\hat{\sigma}_{y}$, respectively. To start, let us assume non-contextuality, denoting the results of the local measurements in $\{ \hat{X} \otimes \hat{X} \}$ by $x_a$ and $x_b$ (subscript $a$ corresponds to the first $\hat{X}$ and subscript $b$ corresponds to the second one), and, similarly, $y_a$ and $y_b$ for the local measurement results of $\{ \hat{Y} \otimes \hat{Y} \}$ \cite{F3}. One expects that the product of the averages of these two operators to be given by:

\begin{equation}
\langle \hat{X} \otimes\hat{X}  \rangle \langle \hat{Y} \otimes\hat{Y}  \rangle = x_a x_b y_a y_b,
\label{clRes}
\end{equation}

\noindent Using exactly the same reasoning when measuring the second set of commuting operators $\{ \hat{X} \otimes \hat{Y}, \hat{Y} \otimes \hat{X}  \}$, we get the following:

\begin{equation}
\langle \hat{X} \otimes\hat{Y}  \rangle \langle \hat{Y} \otimes\hat{X}  \rangle = x_a y_b y_a x_b.
\label{clRes2}
\end{equation}

\noindent If we compare (\ref{clRes}) and (\ref{clRes2}), we come to the following conclusion:

\begin{equation}
\langle \hat{X} \otimes\hat{X}  \rangle \langle \hat{Y} \otimes\hat{Y}  \rangle =\langle \hat{X} \otimes\hat{Y}  \rangle \langle \hat{Y} \otimes\hat{X}  \rangle.
\label{clRes1n2}
\end{equation}

\noindent To reiterate, if we have non-contextual measurements, then (\ref{clRes1n2}) should hold. How can this relation guide us to uncover connections with other non-local properties of the system, such as quantum discord? To start, let us calculate these averages, according to the predictions of quantum mechanics, for a specific state, namely the class of states represented by the Werner state \cite{Werner}:

\begin{equation}
\hat{\rho}= \frac{1-c}{4 }\hat{I}+c\ket{\Psi^{-}}\bra{\Psi^{-}},
\label{Werner}
\end{equation}

\noindent where $c$ is a real parameter, $c \in [-\frac{1}{3},1]$, $\hat{I}$ is the identity operator, and $\ket{\Psi^{-}}=\frac{1}{2} \left( \ket{01} - \ket{10} \right)$ is the singlet state. Notice that this state is a linear combination of a maximally mixed state and a maximally entangled (pure) state. For this class of states, the averages in (\ref{clRes}) and (\ref{clRes2}), using $\langle \hat{O}  \rangle= \mbox{Tr} \{ \rho \hat{O} \}$, where $\mbox{Tr} $ stands for the trace of a matrix, are:

\begin{eqnarray}
\langle \hat{X} \otimes\hat{X}  \rangle \langle \hat{Y} \otimes\hat{Y}  \rangle &=& (-c)(-c) = c^2, \\ \nonumber
\langle \hat{X} \otimes\hat{Y}  \rangle \langle \hat{Y} \otimes\hat{X}  \rangle &=& (0)(0) = 0.
\label{WAve}
\end{eqnarray}

\begin{figure} [b!]
\vspace{-3mm}
\includegraphics[angle=0,width=0.5 \columnwidth]{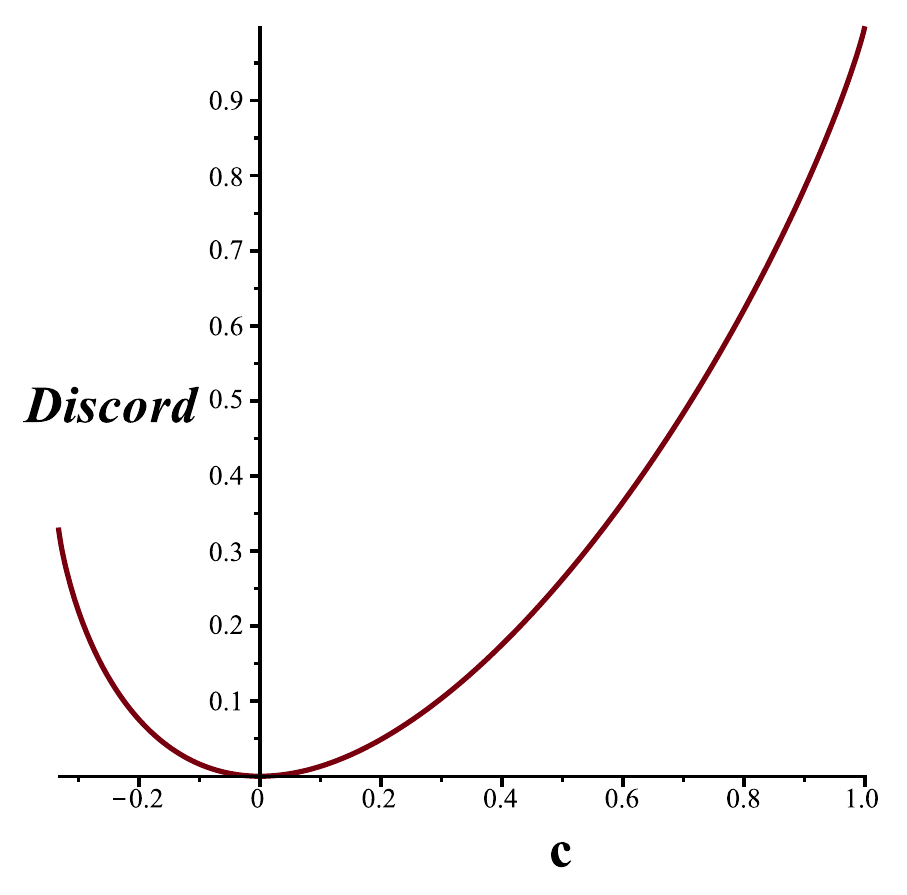}
\vspace{-2mm}
\caption{\textbf{ (Colour Online) Discord versus the parameter $c$ that defines the Werner state.} Notice that discord is only zero at $c=0$ \cite{Luo}.}
\label{fig:WQ}
\end{figure}

\noindent Notice that these two equations are not equal, unless $c=0$; i.e., unless the state is in a maximally mixed state. Well, it looks like it is good news: For Werner states, only when discord is zero (See fig.\ref{fig:WQ}), the system is non-contextual, so perhaps non-contextuality is equivalent to classicality? This example is not enough to make such conclusions because it does not give us enough non-discordant states to analyze, just one. Moreover, for the maximally mixed state, it is not only discord that is zero, but there are no classsical correlations either. We need to find another, more general, two-qubit system that has a \emph{range} of zero-discordant states, not just one, and not just the maximally mixed states (we also want to include classical states in the analysis). We will, therefore, turn our attention now to X-states \cite{YuEberly}, which are states of the form:

\begin{equation}
\hat{\rho}_{X}=
\left(
\begin{array}{cccc} a & 0 & 0 & w  
\\             	   	0 & b & z & 0  
\\									0 & z & c & 0  
\\									w & 0 & 0 & d  
\end{array}
\right ).
\label{rhoX}
\end{equation}

\noindent Without loss of generality, we will assume that all the non-zero elements in (\ref{rhoX}) are real \cite{Xpaper}. First we will consider the set of classical states. A classical state is defined to be a state with zero discord, and it is a state that is diagonal in an orthogonal basis \cite{Modi}. In this work, we are working with the computational basis, which is orthonormal. Therefore, for our X-state in (\ref{rhoX}), a classical state will be one in which $z=w=0$. This can easily be shown to be true by showing that discord is zero for this diagonal state, using equation (23) in \cite{Xpaper}. In fact, discord is non-zero as long as either $z$ or $w$ is finite. Therefore, the classical state is represented by:

\begin{equation}
\hat{\rho}_{classical}=
\left(
\begin{array}{cccc} a & 0 & 0 & 0  
\\             	   	0 & b & 0 & 0  
\\									0 & 0 & c & 0  
\\									0 & 0 & 0 & d  
\end{array}
\right ).
\label{rhoCL}
\end{equation}

\noindent We find that the product averages (for the sets of commuting operators) we discussed earlier for $\hat{\rho}_{classical}$ are given by:

\begin{equation}
\langle \hat{X} \otimes\hat{X}  \rangle \langle \hat{Y} \otimes\hat{Y}  \rangle =\langle \hat{X} \otimes\hat{Y}  \rangle \langle \hat{Y} \otimes\hat{X}  \rangle = 0.
\label{rCLrel}
\end{equation}

\noindent The observation in (\ref{rCLrel}) is consistent with a behaviour in which classicality is equivalent to non-contextuality. Does this mean that if a system has non-zero discord, then the measurements are contextual? It is tempting to say yes, but here is a counterexample. For the following separable state; i.e., a state with no entanglement:

\begin{equation}
\hat{\rho}_{abz}=
\left(
\begin{array}{cccc} \frac{1}{4} + \alpha & 0 & 0 & z  
\\             	   	0 & \frac{1}{4} + \beta & z & 0  
\\									0 & z & \frac{1}{4} - \beta & 0  
\\									z & 0 & 0 & \frac{1}{4} - \alpha 
\end{array}
\right ),
\label{rhoXnc}
\end{equation}

\noindent where, in terms of the original matrix elements, $ \alpha=a-\frac{1}{4}$ and $\beta=b-\frac{1}{4}$ (this re-parametrization makes calculations easier), using equation (23) in \cite{Xpaper}, one can not only show that the X-state (\ref{rhoXnc}), in general, has a non-zero value for discord (although it is under 0.3 and much lower than that for most cases), some exceptions being when $a=b$, $z=0$, or when the system is in a maximally mixed state, but the relations in (\ref{rCLrel}) also hold. That, of course, tells us that for the particular measurement sets we are focusing on here, they can be non-contextual even when there is discord in the system! Therefore, \emph{we cannot equate contextuality to being discordant}.

In conclusion, in this work, we focused on and studied two increasingly popular ways to determine the existence of quantum correlations in a system, namely discord and contextuality. Discord answers the question by capturing the effect of local measurements on the global mutual information function. Whether a system is contextual or not depends on whether assigning local measurement results to operators gives an accurate prediction of the global measurement results of these operators. We showed, using two sets of commuting operators, that for the most general classical two-qubit system; i.e., for the case of zero discord, the system is non-contextual. However, we also showed, using a counterexample, that the opposite is not necessarily true: Having non-zero discord does not imply that the system is contextual. The Kochen-Specker theorem implies that for the two-qubit system, all measurements are contextual, but, taking into account the comparisons made here with discord, what about the maximally mixed states? What about the discordant states that we presented here that are not contextual?

I would like to thank D. F. V. James and  J. H. Eberly for valuable discussions. I would also like to thank an anonymous reviewer for helpful comments on the originally submitted manuscript. This work was supported by the M. Hildred Blewett Fellowship of the American Physical Society, www.aps.org.


\begin{thebibliography}{}
\bibitem{F1} Here, we are referring to quantum particles, which, of course, can also display wave-like properties.
\end{thebibliography}
\end{document}